\documentstyle[epsfig,12pt]{article}
                                                                                       
\newlength{\dinwidth}                                                                                       
\newlength{\dinmargin}                                                                                       
\def\lapproxeq{\lower .7ex\hbox{$\;\stackrel{\textstyle                                                                                       
<}{\sim}\;$}}                                                                                       
\def\gapproxeq{\lower .7ex\hbox{$\;\stackrel{\textstyle                                                                                       
>}{\sim}\;$}}

\def\msb{\overline{\rm MS}}                                                                                       
\def\GeV{{\rm GeV}}

\def\gup{{g\uparrow}}  
\def\gdown{{g\downarrow}}  
\def\asup{{\alpha_S\uparrow\uparrow}}                                                    
\def\asdown{{\alpha_S\downarrow\downarrow}}                                                    
\def\beq{\begin{equation}}                                                    
\def\eeq{\end{equation}}                                                    
\def\bea{\begin{eqnarray}}                                                    
\def\eea{\end{eqnarray}}                                                   
\def\be{\begin{equation}}                                                                                       
\def\ee{\end{equation}}                                                                                       
\def\bea{\begin{eqnarray}}                                                                                       
\def\eea{\end{eqnarray}}                                                                                       
                                                                                       
\begin{document}                                                                                       
\titlepage                                                                                       
\begin{flushright}                                                                                       
DTP/98/52  \\                                                                                       
RAL-TR-1998-061 \\  
August 1998                                                                                       
\end{flushright}                                                                                       
                                                                                       
\begin{center} 
\vspace*{2cm}                                                                                       
{\Large \bf Scheme dependence, leading order and higher twist  \\ [2mm]
studies of MRST partons} \\
\vspace*{1cm}                                                                                       
A.\ D.\ Martin$^a$, R.\ G.\ Roberts$^b$, W.\ J.\ Stirling$^{a,c}$ and R.\ S.\                                              
Thorne$^d$ \\                                                                                       
                                                                                       
\vspace*{0.5cm}                                                                                       
$^a \; $ {\it Department of Physics, University of Durham,                                                                                       
Durham, DH1 3LE }\\                                                                                       
                                                                                       
$^b \; $ {\it Rutherford Appleton Laboratory, Chilton,                                                                                       
Didcot, Oxon, OX11 0QX}\\                                                                                       
                                             
$^c \; $ {\it Department of Mathematical Sciences, University of Durham,                                              
Durham, DH1 3LE} \\                                             
                                                                                       
$^d \; $ {\it Jesus College, University of Oxford, Oxford, OX1 3DW}                                                                                       
\end{center}                                                                                       
                                                                                       
\vspace*{1.5cm}                                                                                       
\begin{abstract}  
We extend a recent global analysis of nucleon parton distributions carried 
out at next-to-leading order (NLO) in the $\msb$ scheme to provide distributions in 
the so-called DIS scheme.  We pay particular attention to the translation of the heavy 
quark distributions in going from the $\msb$ to the DIS scheme.  We repeat 
the global analysis at leading order (LO) and comment on the major 
effects produced when going from LO to NLO.  
Finally we include in the global analysis a freely parameterised form
of possible higher twist contributions 
to make an exploratory study of the size of these effects as a 
function of $x$.
\end{abstract}                                                                                       
                                                                                       
\newpage                                                                                       
                                                                                       
\noindent{\large \bf 1. Introduction}                                                                                       
                                                                                       
We recently published a global analysis (MRST)\footnote{In this paper
we use `MRST' to denote our previous parton
analysis \protect\cite{MRST98} in the
$\msb$ factorisation scheme.} \cite{MRST98} of data in which we
extracted quark and gluon distributions for the proton. The aim of the 
analysis was to constrain the partons by the present data $-$ on deep 
inelastic scattering (DIS) at HERA H1 \cite{H1,ZEUS}, from DIS with
fixed target experiments \cite{NMC,BCDMS,E665,SLAC,CCFR2}, 
and from prompt photon
production (PPP)  \cite{WA70,E706}. In addition the
Drell-Yan measurements \cite{E605,E772} together with the $pp/pn$ asymmetry in
Drell-Yan \cite{NA51,E866} and the asymmetry of the rapidity
distributions of the charged lepton from $W^\pm \rightarrow l^\pm \nu$ decays  
at the Tevatron $p \bar{p}$ collider \cite{CDF} were used in the analysis.
Particular attention was given to the PPP data and the influence they have 
on the determination of the gluon at large $x$. To obtain a satisfactory 
description of the higher energy data \cite{E706} some intrinsic transverse
momentum of the partons needs to be included and by varying the value of 
$\langle k_T \rangle$ within an acceptable range we arrived at a range of 
estimates for the gluon distribution at large $x$. The parton distributions 
corresponding to the extremes of this range were labelled MRST($\gup$) and 
MRST($\gdown$) and these together with the set using the `central' gluon, MRST, 
all used a value of $\alpha_S(M_Z^2) = 0.1175$. In order to reflect the range 
of uncertainty in $\alpha_S(M_Z^2)$ we also provided sets of partons (with 
the central gluon choice) where $\alpha_S(M_Z^2)$ was varied by $\pm
0.005$ and these sets were labelled MRST($\asup$) and MRST($\asdown$).
 
Our analysis included next-to-leading order (NLO) corrections which were computed
in the $\msb$ factorisation 
scheme.  A special feature of the analysis was the new treatment of  
heavy flavour production in DIS, using the procedure developed by two of
us \cite{RTLET,RT}, which describes the threshold behaviour correctly and which is
consistent with the $\msb$ scheme.  The resulting five sets of parton distributions  
were therefore appropriate only to processes evaluated in the $\msb$ scheme.  
 
In this paper we extend our previous analysis to confront a variety of issues.  In  
particular we present\footnote{The {\tt FORTRAN} code for all the parton sets
described in this paper together with those of MRST can be obtained from
{\tt http://durpdg.dur.ac.uk/HEPDATA/PDF},  or by contacting {\tt  
W.J.Stirling@durham.ac.uk}. In addition, because of the slightly complicated  
expressions involving heavy flavours, the packages for computing the structure  
functions from each set are provided there.  A {\tt FORTRAN} routine to compute the charged
current structure functions $F_2$ and $xF_3$ from the MRST parton distributions
has also recently been included.} (i) parton distributions also at NLO but computed in  
the so-called DIS scheme \cite{Diemoz}, (ii) parton distributions resulting from
simply a LO analysis and (iii) parton distributions resulting from a NLO analysis  
which in addition allows for an empirical higher twist contribution. \\ 
 
\noindent{\large \bf 2. Partons in the DIS scheme}          
 
The DIS scheme is simply a device for absorbing the one-loop $\msb$
coefficients into a re-definition of the parton distributions so as to
exactly preserve the NLO value of $F_2$ but using an apparent LO expression.
No new global fitting procedure is involved.  
 
The relation between a quark density in the DIS and $\msb$ schemes is,
at NLO,
\bea 
q^{\rm DIS}_i(x,Q^2) = q^{\msb}_i(x,Q^2)  
&+& \left (\frac{\alpha_S}{4\pi}\right ) 
C^{(1)}_{2,q}(z) \otimes q^{\msb}_i(x/z,Q^2)  \nonumber \\ 
&+& \left (\frac{\alpha_S}{4\pi}\right ) 
C^{(1)}_{g}(z) \otimes g^{\msb}(x/z,Q^2) 
\label{eq:qdis} 
\eea 
where $C^{(1)}_{2,q}(z)$ and $C^{(1)}_{g}(z)$ are the normal $\msb$ 
massless coefficient functions, e.g.  
$$ 
C^{(1)}_{g}(z) = P^{(0)}_{qg}(z)\log ((1-z)/z) +8z(1-z)-1. 
$$ 
The corresponding relation for the gluon is a matter of convention and 
normally one fixes it to maintain the conservation of momentum.   
\bea 
g^{\rm DIS}(x,Q^2) = g^{\msb}(x,Q^2)  
&-& \left (\frac{\alpha_S}{4\pi}\right ) 
C^{(1)}_{2,q}(z) \otimes \Sigma^{\msb}(x/z,Q^2) \nonumber \\ 
&-& 2 n_f \left (\frac{\alpha_S}{4\pi}\right ) 
C^{(1)}_{g}(z) \otimes g^{\msb}(x/z,Q^2) ,
\label{eq:gdis} 
\eea 
where $\Sigma(x,Q^2) = \sum_{i=1}^{n_f}[q_i(x,Q^2) + \bar q_i(x,Q^2)]$, and
where $n_f$  
is the number of light flavours.  
The attraction of the DIS scheme, of course, is that if to LO the $F_2$ 
structure function is given by 
\be 
F_2(x,Q^2) = \sum_i x \;[ a_{2,i} \; q_i(x,Q^2) + \bar a_{2,i} \; 
\bar q_i(x,Q^2) \;] 
\label{eq:f2dis} 
\ee 
then the same expression holds at NLO provided $q_i = q_i^{\rm DIS}$.
However in this scheme it is only the $F_2$ structure function that
has no explicit higher order perturbative corrections. Thus, if at LO
$xF_3$ is given by 
\be  
xF_3(x,Q^2) = \sum_i x [ \; a_{3,i} \; q_i(x,Q^2) + \bar a_{3,i} \;  
\bar q_i(x,Q^2) \; ] ,
\label{eq:f3lo} 
\ee 
then at NLO we have 
\be 
xF_3(x,Q^2) =  \left [\delta (1-z) - \left (\frac{\alpha_S}{4\pi}\right ) 
\frac{8}{3} (1+z) \right ] 
\otimes \sum_i x [ \;a_{3,i} \;q_i^{\rm DIS}(x/z,Q^2) + \bar a_{3,i}  
\; \bar q_i^{\rm DIS}(x/z,Q^2) \; ] .
\label{eq:f3dis}  
\ee 
 
This is the prescription for light favours. For the case of   
heavy quarks then in some scheme which uses $\msb$ evolution for the  
parton distributions, such as that in MRST, the coefficient functions for the  
heavy quarks are mass dependent. Hence we have to decide how best to define a  
DIS scheme for massive quarks. The approach we have adopted is to make exactly 
the same change of definitions for the partons as above, i.e.  
using massless coefficient functions only. Hence the  
heavy quarks in this DIS scheme evolve exactly like the light quarks in  
DIS scheme (in the $\msb$ scheme the evolution starts from zero at $\mu^2=m_H^2$, 
so from Eq.~(\ref{eq:gdis}) we see that in the DIS scheme it starts from a
nonzero value dependent on the gluon distribution). However, the DIS scheme  
coefficient functions for heavy quarks are nontrivial, unlike the case for the light  
quarks.  Expressing first the charm structure function $F_2^c$ in terms of the 
$\msb$ (MRST) charm quark we have 
\be 
F_2^c= \frac{8}{9} C^{(0)}_{2,c}(z,\frac{m^2_c}{Q^2})  
\otimes c^{\msb}(x/z)  
+\frac{8}{9}\left (\frac{\alpha_S}{4\pi}\right ) 
\hat C^{(1)}_{2,g}(z,\frac{m^2_c}{Q^2}) \otimes g^{\msb}(x/z), 
\label{eq:f2cmrst} 
\ee 
and then using Eq.~(\ref{eq:qdis}) to define the DIS charm quark in terms of
the $\msb$ (MRST) charm quark distribution we find        
\be 
F_2^c= \frac{8}{9} C^{(0)}_{2,c}(z,\frac{m^2_c}{Q^2}) \otimes c^{\rm DIS}(x/z)  
+\frac{8}{9}\left (\frac{\alpha_S}{4\pi}\right ) \left [\hat  
C^{(1)}_{2,g}(z,\frac{m^2_c}{Q^2}) -C^{(0)}_{2,c}(z,\frac{m^2_c}{Q^2}) 
\otimes C^{(1)}_{2,g}(z) \right ] \otimes g^{\rm DIS}(x/z), 
\label{eq:f2cdis} 
\ee 
where both coefficient functions depend on $m_c$ (and where in principle  
there is a NLO quark coefficient function which we omit due to its  
insignificance in practice).  All the relevant coefficient functions can be found in  
\cite{RTLET}.  The gluon coefficient function is actually discontinuous at  
$Q^2=m_c^2$, countering the fact that the charm evolution starts from a nonzero  
value, to give a continuous structure function.  The final term above, involving the  
double convolution of two coefficient functions and the gluon distribution, is  
potentially rather complicated, especially since $C^{(0)}_{2,c}$ is defined in terms of  
a convolution itself. However, in practice we find that using $C^{(0)}_{2,c}(z) 
=(1-m_c^2/Q^2)^{1/2}\delta(1-z)$ in this term alone (making the gluon  
coefficient function continuous) gives an extremely good approximation to the true  
result; the contribution from the nonzero charm distribution for $Q^2$ immediately  
above $m_c^2$ being negligible. In the limit $Q^2/m_c^2 \to \infty$, when  
$C^{(0)}_{2,c}(z) \to \delta(1-z)$ and $\hat C^{(1)}_{2,g}(z) \to C^{(1)}_{2,g}(z)$,  
the usual trivial relationship between the structure function  and parton distribution in 
the DIS scheme is regained.  
 
We could alternatively have defined the heavy quark DIS scheme by demanding  
this trivial relationship for coefficient functions. This would lead to very  
complicated, mass-dependent splitting functions in the DIS scheme. Hence we prefer  
to keep all the mass dependence in the coefficient functions, as in the 
$\msb$ scheme, and  
let the partons be the usual DIS partons.  
 
We are therefore able to provide the parton distributions in the DIS scheme analogous to
those in MRST in the $\msb$ scheme \cite{MRST98},
labelled MRSTDIS, MRSTDIS($\gup$),
MRSTDIS($\gdown$), MRSTDIS($\asup$) and MRSTDIS($\asdown$). \\ 
 
\noindent {\large \bf 3.  Leading order parton distributions} 
 
For some purposes, e.g. Monte-Carlo simulation
programs, partons distributions which
attempt to describe the data at the leading order level are preferable.
To obtain such
partons we repeat the global analysis at LO. This means 
that the partons evolve only via the LO DGLAP equations and each process 
is expressed in terms of the partons via LO formulas. The starting
distributions at $Q_0^2 = 1$~GeV$^2$ have the same functional form as in
the NLO analysis \cite{MRST98}.
 
The heavy flavour contributions to the structure function $F_2$ are computed to LO  
which means that for $Q^2 < m_c^2$ we have 
\beq 
F_2^c (x,Q^2) = \frac{8}{9} \left ( \frac{\alpha_S}{4\pi} \right )  
C_g^{(1) {\rm FF}}(z,m_c^2/Q^2) \otimes g(x/z,Q^2), 
\label{eq:fflo} 
\eeq  
while for $Q^2 > m_c^2$ 
\beq 
F_2^c (x,Q^2) = F_2^c (x,Q^2 = m_c^2) 
+ \frac{8}{9} C_c^{(0)} (z,m_c^2/Q^2) \otimes c(x/z,Q^2). 
\label{eq:vflo} 
\eeq 
Again, the coefficient functions in Eqs.~(\ref{eq:fflo},\ref{eq:vflo}) 
can be found in \cite{RTLET}. 
 
There is a significant difference in the sizes of the LO and NLO 
gluon distributions at large $x$  
which reflect the importance of the NLO corrections to the prompt  
photon production process (PPP) in that region. Thus to get the same cross section 
the LO gluon has to be typically greater than the NLO gluon by about 40-50\%  
for $x = 0.3-0.45$. The LO gluon is also larger than the NLO distribution at 
small $x$, this being required for a good description of the small-$x$ HERA 
data, and reflecting the importance of the NLO corrections to the quark  
evolution at small $x$. This leads to a quite different prediction for the  
longitudinal structure function (which has to be taken into account when  
obtaining the values of $F_2(x,Q^2)$ from measurements of the cross section). 
For example the LO value of $F_L$ is twice the NLO value at $x=10^{-4}$ and  
$Q^2=4$ GeV$^2$.  
This again illustrates the importance of a precise measurement of $F_L$
at small $x$ as a test of higher order perturbative corrections.
We note also that the Drell-Yan LO cross section requires a phenomenological 
$K$--factor of the order of
$+30\%$ in order to get acceptable agreement with experiment.
 
To obtain a reasonable description of the DIS data we require a larger value of 
$\alpha_S(M_Z^2)$ than in the NLO case, but too large a value spoils the
simultaneous 
description of $F_2$, $F_2^c$ and the PPP data. Using the simple scale choice of 
$\mu^2=Q^2$ we find a value of $\Lambda_{\rm LO}$(4 flavours) = 174 MeV; a  
satisfactory compromise which implies a value $\alpha_S(M_Z^2) = 0.125$. We  
define our `central' LO solution -- MRSTLO -- with this value of $\alpha_S$ and with
a gluon constrained by the PPP data as before.\footnote{The same value
$\langle k_T \rangle = 0.4\; \GeV ( 0.92\; \GeV)$ for WA70 (E706--530~GeV)
is used as in Ref.~\protect\cite{MRST98}.} The MRSTLO($\asup$) and
MRSTLO($\asdown$) solutions again correspond to varying this central value 
of $\alpha_S(M_Z^2)$ by $\pm 0.005$.
   
\begin{figure}[H] 
\begin{center}      
\epsfig{figure=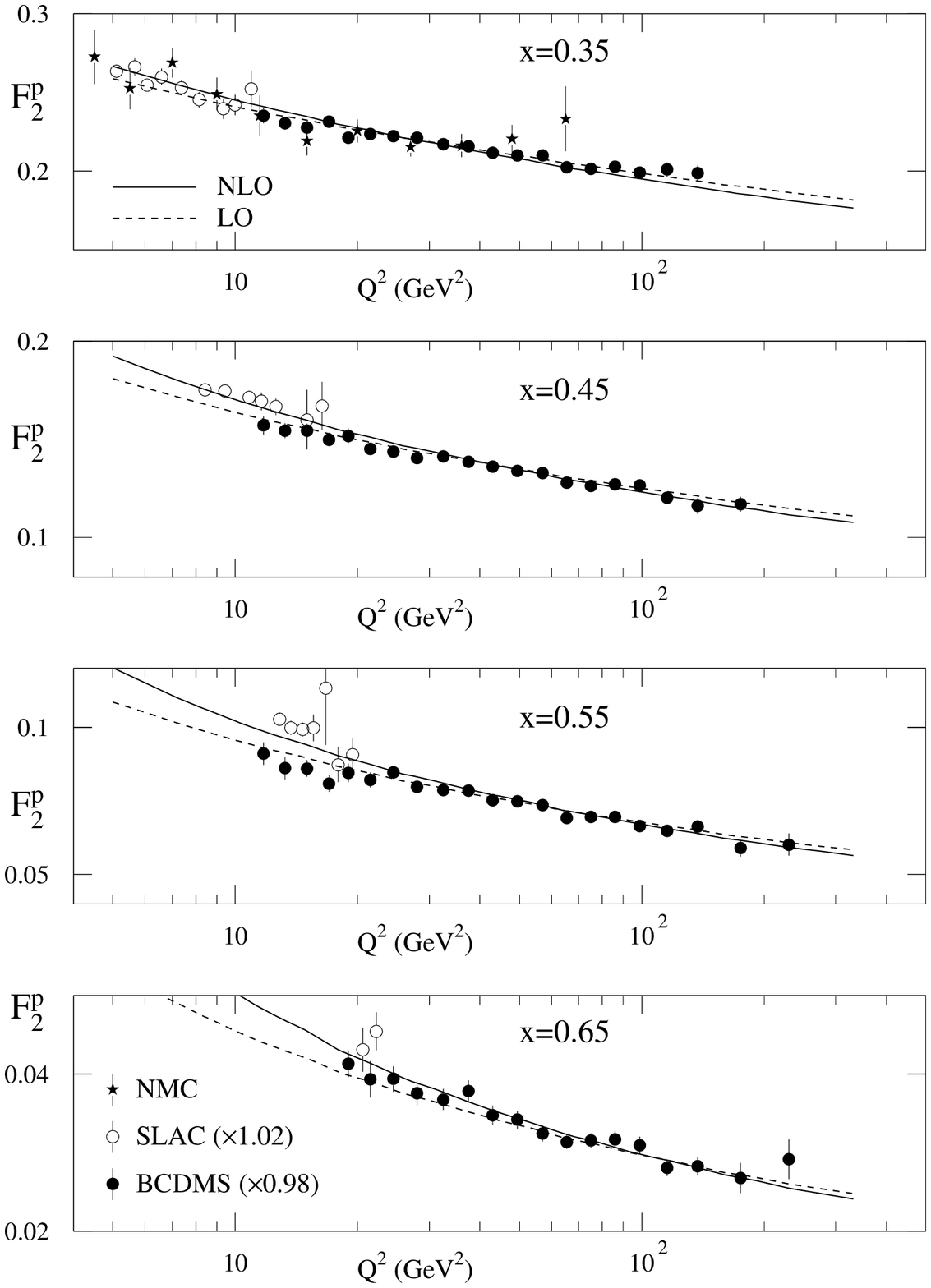,height=20cm}
\end{center}     
\caption{Comparison of large $x$ data from BCDMS, NMC and SLAC with 
LO and NLO fits of MRST.}
\label{fig:fig1} 
\end{figure}

Overall, the quality of the LO and NLO fits are comparable, see
Table 1, with two exceptions.
\begin{table}[htb]
\begin{center}
\begin{tabular}{|cccc|}\hline                                                                               
Data set          & No. of   & MRST & MRSTLO\\                                                                               
                  & data pts &      &      \\ \hline                                                                               
H1 $ep$           & 221      & 164  & 159  \\                                                                               
ZEUS $ep$         & 204      & 269  & 288  \\                                                                               
BCDMS $\mu p$     & 174      & 248  & 171  \\                                                                               
NMC $\mu p$       & 130      & 141  & 144  \\                                                                               
NMC $\mu d$       & 130      & 101  & 115  \\                                                                               
SLAC $ep$         & 70       & 119  & 188  \\                                                                               
E665 $\mu p$      & 53       & 59   & 52   \\                                                                               
E665 $\mu d$      & 53       & 61   & 62   \\                                                                               
CCFR $F_2^{\nu N}$& 66       & 93   & 79   \\                                                                               
CCFR $F_3^{\nu N}$& 66       & 68   & 83   \\ \hline                                                                              
\end{tabular}                                                                               
\caption{The $\chi^2$ values for the DIS data included in the (NLO--$\msb$)
MRST fit of Ref.~\protect\cite{MRST98} and the corresponding LO fit.}
\end{center} 
\label{tab1}
\end{table}
The SLAC data (which cover a region\footnote{Only data for $x <
0.7$ are included in these fits -- see the discussion in the next
section.}
of large $x$ and
relatively low $Q^2$) strongly prefer the NLO corrections $-$ a fact established a  
long time ago \cite{dr}. The effect of these corrections at large $x$ is equivalent to
using a LO description where the value of $\alpha_S$ increases with $n$, the moment  
of the structure function \cite{bbdm,dr,bllm}.  While this effect helps the description
of the SLAC data, the BCDMS data actually do not favour this trend. Consequently,  
despite the relatively large value of $\alpha_S$, the absence of NLO corrections can  
approximately mimic at large $x$ a NLO fit with a lower value of $\alpha_S$. Thus  
the BCDMS data are surprisingly well described by our LO global fit which can 
be compared with the MRST(NLO) description in Fig.~\ref{fig:fig1}.  \\
 
\noindent {\large \bf 4.  Global analysis including higher twist terms} 
            
In MRST we showed that our NLO fit slightly underestimated the slope
$dF_2/d\log Q^2$ for the NMC data since the low $Q^2$ data tended to lie 
systematically below the fit. Clearly the inclusion of a negative $1/Q^2$ 
contribution is bound to improve the quality of the description in this
$x$ range $ 0.02 - 0.1$.  Higher twist terms have always been expected to play
an important role at very large $x$ $-$ indeed we found in the past \cite{lowq2}
that the SLAC data for $x > 0.7$ were dominated by power corrections and 
for that reason were excluded in MRS leading twist analyses. In those 
analyses, including MRST, we also imposed a lower $W^2$ cut on all data of
10~GeV$^2$ to reduce the effect of unknown higher twist contributions. Now
that we are allowing such terms, in this section we relax the constraint so that only  
data for $W^2 < 4$~GeV$^2$ are removed.  In addition we have lowered the
$Q^2$ cut on data included in the fit from 2 to 1.2~GeV$^2$.
   
At very small $x$ we are also interested in examining whether $1/Q^2$  
corrections may be important. Some attention has been paid to the observation 
that at HERA the slope $dF_2/d\log Q^2$ appears to `flatten' off  
around $x=10^{-4}$ \cite{caldwell} in contrast to the naive DGLAP
expectation.
In MRST this was attributed to a `valence-like' behaviour
of the gluon at the starting scale $Q_0^2 = 1$~GeV$^2$.  That is for $x$
below $10^{-3}$ the gluon is suppressed and consequently, since for the HERA data  
$x$ and $Q^2$ are strongly correlated, this leads to a leveling off of 
the slope at very low values of $x$ and $Q^2$. However this `valence-like' 
behaviour of the gluon may be an artifact, reflecting some dynamics 
other than DGLAP, and one candidate is a positive higher twist contribution
which is relevant only at small $x$. 
 
We assume a very simple parameterisation of the higher twist contribution
to the DIS structure function,
\beq 
F_2^{\rm HT}(x,Q^2) = F_2^{\rm LT}(x,Q^2)\;\left ( 1+ \frac{D_2(x)}{Q^2}
 \right) ,
\label{eq:f2ht} 
\eeq     
where the leading twist NLO structure function $F_2^{\rm LT}$ is treated exactly 
as in MRST. Applying the same overall parameterisation independent of
target or beam is probably an over-simplification but may have some 
justification if the higher twist contribution can be derived through renormalon
dynamics \cite{renormalon}.
  
We parameterise the coefficient $D_2(x)$ by a constant over different bins 
in $x$ chosen to emphasize aspects of different datasets. 
We perform fits 
using the `central' gluon type of solution. In general we find that for  
$x < 0.5$ the resulting correction is small and negative but beyond 0.6 
large and positive. The values of $D_2(x)$ obtained from the fit
in each $x$ bin are shown in Table 2.
\begin{table}[htb]
\begin{center} 
\begin{tabular}{|cc|}\hline 
  $x$ & $D_2(x)$ (GeV$^2$) \\ \hline 
 0 -- 0.0005 & 0.0147 \\
 0.0005 -- 0.005 & 0.0217 \\
 0.005 -- 0.01 & $-$0.0299 \\
 0.01 -- 0.06 & $-$0.0382 \\
 0.06 -- 0.1 & $-$0.0335 \\
 0.1 -- 0.2 & $-$0.121 \\
 0.2 -- 0.3 & $-$0.190 \\
 0.3 -- 0.4 & $-$0.242 \\
 0.4 -- 0.5 & $-$0.141 \\
 0.5 -- 0.6 & 0.248 \\
 0.6 -- 0.7 & 1.458 \\
 0.7 -- 0.8 & 4.838 \\
 0.8 -- 0.9 & 16.06  \\ \hline
\end{tabular} 
\caption{The values of the higher twist coefficient $D_2(x)$
of Eq.~(\ref{eq:f2ht}) (in GeV$^2$) versus $x$.}
\end{center} 
\label{tab2}
\end{table}
 
Looking at the overall improvement of the resulting fit we note no real 
difference in the comparison with the HERA data, indeed the higher twist
contributions chosen for $x<0.01$ are very small.  
We have tried fits where 
the starting gluon at $Q_0^2 = 1$~GeV$^2$ was forced to be flat, or even singular, 
as $x \rightarrow 0$ to see if the effect of the valence-like gluon could 
be described instead by a significant positive higher twist term. In each case 
the quality of the resulting fit was (far) worse. So there appears to be no  
preference for a description in terms of a `conventional' gluon 
at low $Q^2$ with additional power corrections, at least within our  
admittedly simple parameterisation. 
 
\begin{figure}[H] 
\begin{center}      
\epsfig{figure=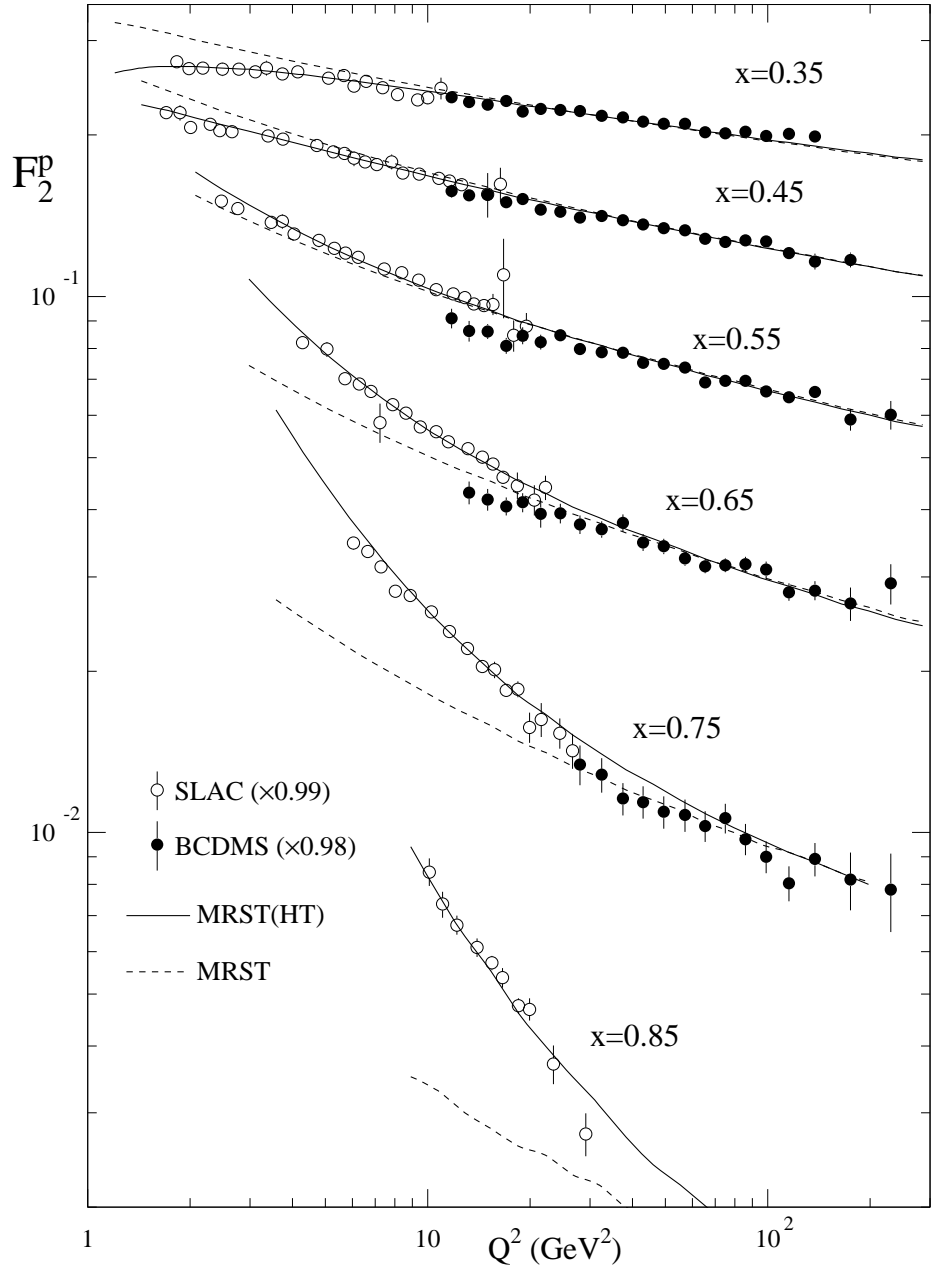,height=20cm}
\end{center}     
\caption{Comparison of very large $x$ data from BCDMS and SLAC with the  
standard NLO fit of MRST and the higher twist fit MRST(HT).}
\label{fig:fig2} 
\end{figure}

At intermediate $x$ values ($0.01<x<0.5$) there is definitely a preference  
for some negative 
$1/Q^2$ contribution as we expect from the NMC data. With the relaxing of 
the $Q^2$ cut, the number of NMC $F_2^p$ points increases from 130 to 155 
but the $\chi ^2$ stays close to 140 with the inclusion of the higher twist 
term. The most significant improvement in the quality of the fit is for the  
data for $x>0.1$ where the higher twist corrections are largest. 
The most dramatic difference is the description of the SLAC data.  
In Fig.~\ref{fig:fig2} we illustrate the improvement at very large $x$ due
to the addition of the arbitrary higher twist contribution. From 
this figure and from the values of the coefficient $D_2(x)$
in Table~2 we see that the very large $x$ SLAC data require a large positive
$1/Q^2$ correction. We have not attempted to separate the part arising
from target mass corrections but a recent analysis \cite{ybf} suggests
a significant fraction can be accounted for in this way.
 
\begin{figure}[H] 
\begin{center}      
\epsfig{figure=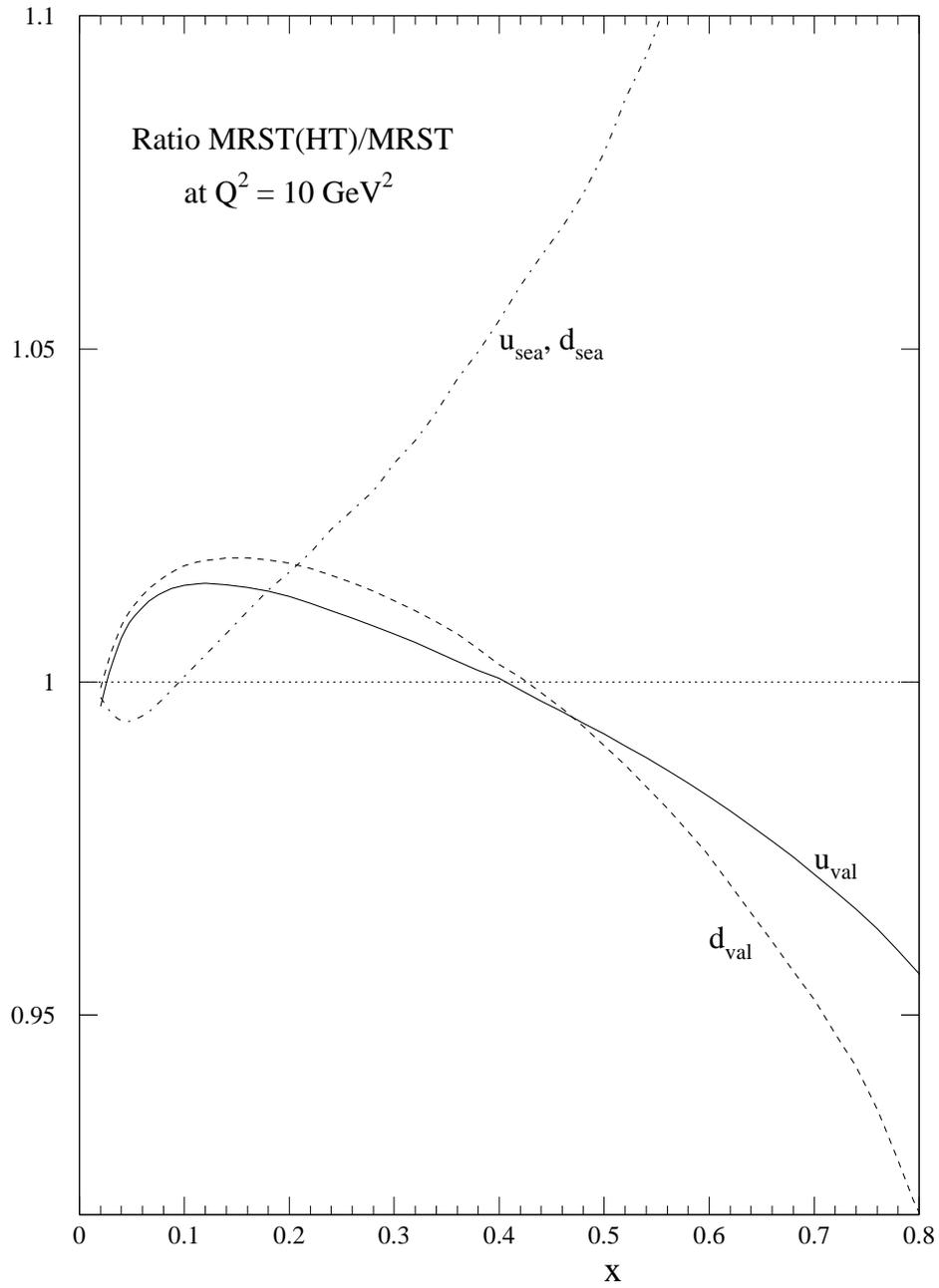,height=20cm}
\end{center}     
\caption{Ratios of $u$ and $d$ partons in the two sets MRST(HT) and MRST.} 
\label{fig:fig3} 
\end{figure}

We may regard the difference between this new set of partons --
MRST(HT) --
and MRST as one measure of uncertainty in our parton sets.  
In Fig.~\ref{fig:fig3} we show the ratio of the $u$ and $d$ partons
at $Q^2 = 10$ GeV$^2$ between the  
two sets $-$ the differences being much smaller for other partons. We see 
that the differences remain less than about 1-2\% 
except where the individual distributions start to become really 
quite small.  At large $x$, $x > 0.6$, the large differences for the valence distributions 
are not surprising in view of the large higher twist correction as 
$x \rightarrow 1$. As $Q^2$ increases the ratios remain rather constant 
so that Fig.~\ref{fig:fig3} is a reliable indication of the uncertainty on the
$u$ and $d$ partons at all $Q^2$ arising from possible higher twist 
contributions. \\ 
 
\noindent {\large \bf 5. Conclusions} 
  
We have presented an extension of our previous global analysis to provide  
alternative parton distributions. In particular we have obtained the 
analogues of the partons of MRST but (a) in the DIS NLO scheme as
opposed to the $\msb$ scheme and (b) by repeating the global analysis at LO. In this 
latter case the partons are consistent with all processes considered being evaluated to 
LO.  Finally, we present a set of partons obtained from a global analysis in which an 
empirical universal higher twist contribution is included, which is freely 
parameterised 
as a function of $x$.  This empirical higher twist component is found to be 
surprisingly small in the HERA small $x$ domain.  On the other hand it is interesting 
to see that the higher twist fit is similar at high $x$ to that expected from a 
renormalon  
approach for the nonsinglet structure function \cite{renormalon}.

\newpage  
          
\newpage  
          
\newpage  
          
\end{document}